\documentclass[
reprint,            
twocolumn,          
aps,
prd,
amsfonts,
amsmath,
amssymb,
superscriptaddress, 
floatfix,
]{revtex4-2}

\usepackage[T1]{fontenc}

\usepackage{float}
\usepackage{graphicx}   
\usepackage{upgreek}    
\usepackage{comment}
\usepackage{color}
\usepackage[
colorlinks=true,        
citecolor=blue,         
linkcolor=blue,         
urlcolor=blue           
]{hyperref}             
\usepackage{orcidlink}

\usepackage{siunitx}

\newcommand{\DMunits}{\,pc\,cm$^{-3}$}
\newcommand{\RMunits}{\,rad\,m$^{-2}$}
\newcommand{\RRMunits}{\,rad\,m$^{-3}$}

\newcommand{\araa}{Ann. Rev. Astro. Astrphys.}
\newcommand{\mnras}{Mon. Not. R. Astron. Soc.}
\newcommand{\pasa}{Publ. Astron. Soc. Aust.}
\newcommand{\aapr}{Astron. Astrophys. Rev.}
\newcommand{\apjl}{Astrophys. J. Lett.}
\newcommand{\jplp}{J. Plasma Phys.}
\newcommand{\sovast}{Sov. Astron.}
\newcommand{\sovphyu}{Sov. Phys. Usp.}
\newcommand{\apss}{Astrophys. Space Sci.}
\newcommand{\tnsan}{Transient Name Server AstroNote}

\newcommand{\Swin}{\affiliation{Centre for Astrophysics and Supercomputing, Swinburne University of Technology, P.O. Box 218, Hawthorn, VIC 3122, Australia}}
\newcommand{\weizmann}{\affiliation{Department of Particle Physics and Astrophysics, Weizmann Institute of Science, Rehovot 7610001, Israel}}

\newcommand{\ATNF}{\affiliation{Australia Telescope National Facility, CSIRO, Space and Astronomy, PO Box 76, Epping, NSW 1710, Australia}}
\newcommand{\UCSC}{\affiliation{University of California, Santa Cruz, 1156 High St., Santa Cruz, CA 95064, USA}}
\newcommand{\IPMU}{\affiliation{Kavli Institute for the Physics and Mathematics of the Universe,
5-1-5 Kashiwanoha, Kashiwa, 277-8583, Japan}}
\newcommand{\NAOJ}{\affiliation{Division of Science, National Astronomical Observatory of Japan,
2-21-1 Osawa, Mitaka, Tokyo 181-8588, Japan}}

\begin{document}

\title{Propagation of a fast radio burst through a birefringent relativistic plasma}

\author{Pravir Kumar\,\orcidlink{0000-0003-1913-3092}}\email{pravir.kumar@weizmann.ac.il} \Swin \weizmann
\author{Ryan M. Shannon\,\orcidlink{0000-0002-7285-6348}}\email{rshannon@swin.edu.au}\Swin 
\author{Marcus~E.~Lower\,\orcidlink{0000-0001-9208-0009}}\email{marcus.lower@csiro.au}\ATNF 
\author{Adam T. Deller\,\orcidlink{0000-0001-9434-3837}} \Swin
\author{J. Xavier Prochaska\,\orcidlink{0000-0002-7738-6875}} \UCSC \IPMU \NAOJ

\date{\today}

\begin{abstract}
The study of impulsive astrophysical radio emission makes it possible to probe the intervening plasma between the emission source and the Earth. In cold electron-ion plasmas, the circular propagating wave modes primarily alter the linear polarization plane that scales with the inverse-square of the emission frequency. In relativistic plasmas, the wave modes are elliptically polarized, and it is possible to convert linearly polarized emission into circular and vice versa.
Fast radio bursts (FRBs) enable the study of not only the electron-ion plasma of the intergalactic medium but potentially the extreme magneto-ionic medium in which these intense pulses are produced. Here we report on the polarimetric analysis of a repeat burst from the FRB\,20201124A source. The burst displayed a significant frequency-dependent circularly polarized component, unlike other bursts from this source or any other FRB found to date. We model the frequency dependence of the circular polarization using a phenomenological generalized Faraday rotation framework. We find that the observed circular polarization in the burst can be adequately explained by the propagation of an initially linearly polarised burst signal through a relativistic plasma along the line of sight from the progenitor.
\end{abstract}

\maketitle

\emph{Introduction ---} Electromagnetic waves in plasmas have multiple possible propagating wave modes, contrary to a single mode in the vacuum. If the plasma is magnetized, these modes can exhibit a variety of dispersion relations, meaning they have different phase velocities dependent on the wave frequency and the angle of propagation with the background magnetic field. While the propagation delay between these modes is small, it results in a frequency-dependent transformation in the polarimetric properties of the propagating radiation.
In a plasma dominated by cold electrons, the natural wave modes are approximated to be circularly polarized, resulting in a wavelength-dependent change in the plane of linear polarization (LP). This primary propagation effect is the conventional Faraday rotation (FR) \citep{Faraday:1846}, often used to study the magnetic field strength of the cold, magnetized plasma of the interstellar medium (ISM). However, in a plasma dominated by relativistic particles, the modes are not necessarily circularly polarized, and magnetic birefringence can lead to the conversion between the linear and circular polarization (CP) components as a function of frequency \citep{Sazonov:1969, Pacholczyk:1973, Jones:1977, Kennett:1998}. This effect of a generalized form of Faraday rotation (GFR; occasionally referred to as Faraday conversion or Faraday repolarization) has been used to explain the observed CP in the radio spectra of compact astrophysical objects \cite{Wardle:1998, Macquart:2002}.

Astrophysical plasma provides a unique opportunity to study electrodynamics in extreme environments impossible to create in a terrestrial lab. First discovered over a decade ago, fast radio bursts (FRBs) present a new probe of astrophysical plasmas and the highly energetic radiation processes originating there \cite{Lorimer:2007}. These ms-duration transients have been found to originate from galaxies as distant as redshift $z=0.66$ (a luminosity distance of $d_L=4.08$\,Gpc) \cite{Ravi:2019}, with their emission detected only at radio wavelengths \cite{Petroff:2019}. Their extragalactic origin allows us to probe the cold plasma within the galaxies and the intergalactic medium (IGM) along the line of sight (LOS) \citep{Bhandari:2021}. A sample of localized bursts has been used to detect the entirety of the baryon density of the local Universe \citep{Macquart:2020}. The bursts represent the most luminous pulses of radio emission known, with inferred radio luminosities a factor of $\sim10^{12}$ larger than the Galactic radio pulsars, assuming the same beaming angles \cite{Cordes:2019}. With these extremities, the FRB progenitors and their immediate environment most likely contain highly magneto-ionized relativistic plasma. Interactions between the propagating EM radiation and the surrounding plasma can imprint various wavelength-dependent effects, enabling the study of relativistic plasma dispersion relations and their non-linear deviations \citep{Arons:1986, Melrose:1991}.

It is possible to study in great detail the spectro-temporal-polarimetric properties of the $\approx$4\% of FRB sources found to repeat \citep{Petroff:2022}. Most repeating FRB sources display a pulse-averaged LP $\approx$~90--100\% and non-varying LP position angle (PA) across the burst profile \citep{Caleb:2021}. These features suggest a similar emission mechanism for repeating FRBs and are commonly interpreted as relativistic shocks involving synchrotron masers far from the progenitor \citep{Metzger:2019, Beloborodov:2020}. However, recent discoveries of PA variations \citep{Luo:2020, Kumar:2022} and CP \citep{Hilmarsson:2021_circular, Kumar:2022} in a few repeating FRBs have complicated this picture in favor of a magnetospheric origin. Additionally, some repeating FRB sources have been found to be embedded in extremely dynamic magneto-ionic environments \citep{Michilli:2018, Niu:2021}. FRB\,20121102A \citep{Yaron:2020}, the first known repeating FRB source, emits highly linearly polarized bursts with an extraordinarily high and variable Faraday RM~$\approx$~$10^5$\,\RMunits, indicating a dynamic magnetic environment \citep{Michilli:2018, Hilmarsson:2021}. Targeted studies of the variations in burst properties such as the dispersion measure (DM), RM, PA, and scattering levels can be used to probe the local environment of the progenitor.

\begin{figure}
\centering
\includegraphics[width=1\columnwidth]{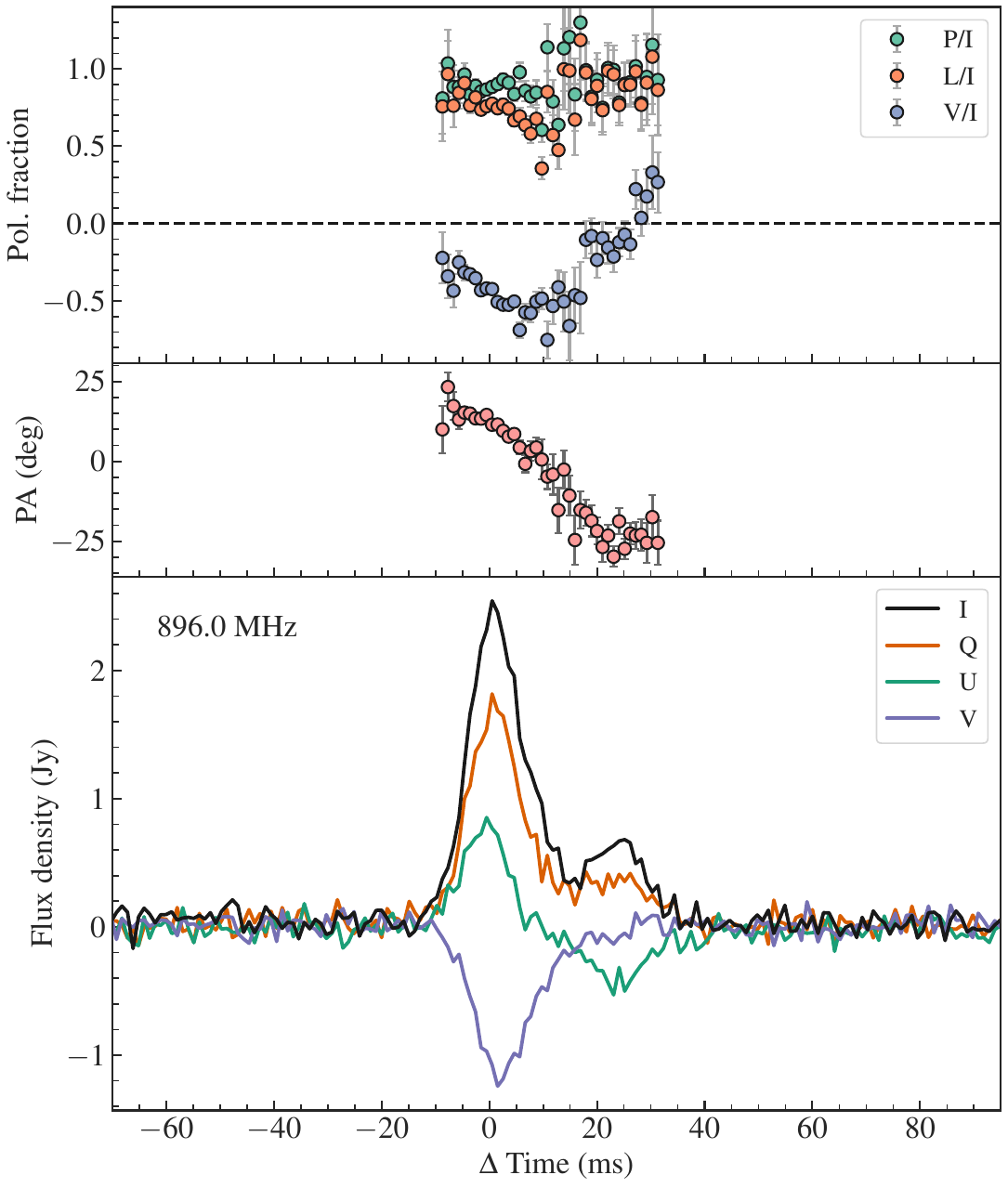} 
\caption{Frequency-averaged profile of the Parkes-detected repeat burst FRB\,20210405H. The top panel shows the degree of polarization components (total ($P/I$), linear ($L/I$), and circular ($V/I$)). The middle and bottom panels show the LP position angle (PA) and the four Stokes parameters. The data are corrected for the first sub-pulse best-fit RM of $-619$~\RMunits~at a reference frequency of 896 MHz.}
\label{fig:burst}
\end{figure}

\textit{FRB\,20210405H ---} A bright (isotropic energy E$_{\rm iso} \sim 10^{41}$~erg) repeat burst FRB\,20210405H was detected on 2021 April 5 during monitoring of the FRB\,20201124A source with the CSIRO Parkes 64-m radio telescope (also known as \textit{Murriyang}) \citep{Kumar:2022}. The DM of this individual burst is estimated to be $418\pm1$\DMunits, which is in excess of $\sim$\,5\DMunits\, from the statistical average ($413\pm1$\DMunits) determined using a sample of bright bursts from this source \citep{Kumar:2022}. The spectral occupancy of the burst is constrained between 704--1088 MHz with no signal detected between 1088--4032\,MHz. The burst consists of two distinct sub-pulses (C1 and C2) separated in time. The burst has a high (90\%) polarization fraction with a likewise high fraction of LP (74\% for C1 and 93\% for C2). A significant amount of CP (47\%) is detected in sub-pulse C1, never seen before in bursts from a repeating FRB source \citep{Kumar:2022}. Figure \ref{fig:burst} shows the frequency-averaged polarization profile of the burst for all four Stokes parameters and the degree of polarization components as a function of time.

\textit{Circular Polarization ---} The production of large fractions of CP in astrophysical radio emission is a long withstanding problem \citep{Sazonov:1973}. The most prominent explanation explored is for it to be intrinsic to the emission. Even though some amount of CP is expected from intrinsic synchrotron radiation \citep{Legg:1968, Melrose:1971, Sazonov:1972}, the predicted level, however, is inadequate to explain the observations, and the expected degree of CP $\propto \lambda^{1/2}$ is generally not observed \citep{Saikia:1988}. For FRB\,20201124A, the majority of detected bursts have little to no CP, with a sub-sample exhibiting varying degrees of CP \citep{Hilmarsson:2021_circular, Kumar:2022}. Thus, if the high CP in FRB\,20210405H was entirely intrinsic to the progenitor, it cannot by itself explain the non-detection in most of the bursts. The inability of prevalent emission models to explain such a high degree of CP through an intrinsic mechanism suggests it is likely the result of a propagation effect in which LP has been converted to CP \citep{Jones:1977, Kennett:1998}. Multiple underlying conditions related to the magnetic field strength and the electron density can be responsible for such a conversion. Nonetheless, the critical requirement is that the EM wave propagates through a strongly magnetized region or a mildly to highly relativistic medium (even with a relatively weak magnetic field strength).

In a non-relativistic cold plasma, the circularly polarized wave propagation modes result in only regular Faraday rotation between the two orthogonal LP components. However, propagation through several magnetic field reversals in a highly magnetized region can allow a partial conversion from LP to CP \citep{Melrose:2010, Gruzinov:2019}. A key prediction from this model is the presence of quasi-periodic oscillations in Stokes V as a function of $\lambda^2$, with the number of quasi-periods corresponding to the number of magnetic-field reversals along the LOS \citep{Gruzinov:2019}. We do not see such oscillatory features in the observed spectra for the Parkes FRB. Thus, we rule out the prospect of a GFR induced by a non-relativistic medium in our case.

Scintillation in the magnetized, cold plasma of ISM is another way to generate CP \citep{Macquart:2000}. This method does not require an initial source of polarized radiation, and a birefringent screen along the LOS can stochastically generate the CP. Another model proposes that a magnetized, scintillating screen may result in the multi-path propagation of radiation and induce a significant CP \citep{Beniamini:2022}. However, at least for this burst, these models cannot explain the contrasting polarimetric properties of the two sub-pulses of the burst, nor the significant sweep in PA observed across sub-pulse C1. Rather, these properties are most likely due to the intrinsic emission mechanism and magnetospheric geometry. Similar behavior is seen in observations of radio pulsars and magnetars \citep{Noutsos:2009, Dai:2019, Ilie:2019}.

The observed high degree of CP and its strong frequency dependence in the Parkes-detected repeat burst thus suggest propagation-induced transmission effects. Such behavior can arise if the burst has undergone a form of GFR after propagating through a relativistic medium with elliptically polarized wave modes \citep{Kennett:1998}.

\textit{Generalised Faraday rotation ---} To consider the effects of Faraday conversion in an elliptically polarized propagating medium, we model the rotation of polarization vector using a phenomenological GFR model by projecting the polarization data onto the Poincar\'{e} sphere \citep{Lower:2021}. The polarization vector represents an arbitrary elliptical mode with its tip on the surface of the sphere. The normalized vector can then be expressed in Cartesian coordinates as $\mathbf{P(\nu)} = [Q(\nu), U(\nu), V(\nu)]/P$, where $P = \sqrt{Q^2 + U^2 + V^2}$ is the total polarization and $Q$, $U$ and $V$ are the Stokes parameters. The LP position angle $\Psi = \frac{1}{2}\tan^{-1}(U/Q)$ and the ellipticity angle $\chi = \frac{1}{2}\tan^{-1}(V/\sqrt{Q^2 + U^2})$ is the latitude on the Poincar\'{e} sphere. The natural wave mode of the birefringent medium is represented as a diagonal axis on the sphere. As the radiation propagates through the medium, birefringence results in rotation of $\mathbf{P(\nu)}$ in a plane perpendicular to the mode's diagonal axis.

In the case of regular FR, the plasma's natural wave modes are circularly polarized, and the mode diagonal is aligned with the vertical $V$ axis connecting the poles of the sphere. Thus, $\mathbf{P(\nu)}$ rotates at a constant latitude ($\chi$) parallel to the equatorial plane depending upon the incident polarization state. This corresponds to a wavelength-dependent change in the PA given by
\begin{equation}\label{eqn:psi_fr}
    \Psi_{\rm FR}(\lambda) = \Psi_{0} + {\rm RM} (\lambda^{2} - \lambda_{c}^{2}),
\end{equation}
where $\Psi_{0}$ is the PA at a reference wavelength $\lambda_{c}$ and
\begin{align}\label{eqn:rm_fr} 
{\rm RM} &= \nonumber \frac{e^3}{8 \pi^2 \varepsilon_0 m_e^2 c^3} \int d\ell n_e B_\parallel \\ 
         &= 0.81\,L\,\langle n_e B_\parallel \rangle\,{\rm rad\,m^{-2}},
\end{align}
with the path length $L$ in parsecs, magnetic field $B$ in $\upmu$G and the cold electron number density $n_e$ in cm$^{-3}$. Whereas in the case of GFR, the polarization of the plasma wave modes can be arbitrary and not necessarily circular. In such a medium, the mode diagonal could be along any arbitrary axis on the Poincar\'{e} sphere, with the polar angle reflecting the ellipticity of the wave modes. The wavelength-dependent rotation along the diagonal axis causes  $\chi$ to vary, thus converting LP into CP. Using an arbitrary parameter $\alpha$ for the wavelength dependence and an analogous ``generalized rotation measure'' (GRM) \citep{Kennett:1998, Lower:2021}, Equation~(\ref{eqn:psi_fr}) can be expressed as
\begin{equation}\label{eqn:grm}
    \Psi_{\rm GFR}(\lambda) = \Psi_{0} + {\rm GRM} (\lambda^{\alpha} - \lambda_{c}^{\alpha}),
\end{equation}
The effects of non-circular modes in the propagating medium are emulated by rotating the polarization vector about the Stokes $U$ and $V$ axes by angles $\vartheta$ and $\varphi$ respectively, as
\begin{equation}
\mathbf{P}_{\rm GFR}(\lambda) = \mathbf{R}_{\vartheta} \cdot \mathbf{R}_{\varphi} \cdot \mathbf{P}(\lambda).
\end{equation}
where
\begin{equation}
    \mathbf{R}_{\vartheta} = 
    \begin{bmatrix}
    \cos(\vartheta) & 0 & \sin(\vartheta)\\
    0 & 1 & 0\\
    -\sin(\vartheta) & 0 & \cos(\vartheta)
    \end{bmatrix},
\end{equation}
and
\begin{equation}
    \mathbf{R}_{\varphi} = 
    \begin{bmatrix}
    \cos(\varphi) & -\sin(\varphi) & 0\\
    \sin(\varphi) & \cos(\varphi) & 0\\
    0 & 0 & 1
    \end{bmatrix},
\end{equation}
Here the ellipticity of the wave modes are parametrized by the orientation of the mode diagonal on the sphere ($\vartheta = 0^{\circ}$ for circular and $90^{\circ}$ for linear).

\textit{Analysis ---} The FRB Stokes spectra include contributions from the signal interaction with the cold ionized ISM, IGM, and a likely relativistic medium near the emission region. Ideally, we would perform a joint FR-GFR fit to simultaneously account for all such contributions \citep{Lower:2021}. However, in our case, a joint fit is neither attainable nor constraining due to the limited bandwidth of the burst. Hence, we proceed by first constraining the regular FR contributions in the sub-pulses and then modeling the effects of GFR in the de-Faraday rotated residuals.

We model the regular FR contribution by directly fitting the rotation in Stokes $Q$ and $U$ spectra of the two sub-pulses. We obtain median \textit{a-posteriori} RM values of $-619\pm1$ and $-613\pm3$\RMunits~for C1 and C2, respectively. The de-Faraday rotated LP component of sub-pulse C1 exhibits a residual frequency-dependent variation, indicating a deviation from the standard $\lambda^{2}$ expected from a regular FR in the intervening medium. The frequency-dependent deviation establishes further evidence of a partial conversion between the polarization components. However, a critical question is whether some of the GFR effects are also getting absorbed in the estimated RM of C1. A piece of compelling evidence is the RM discrepancy between both the sub-pulses, which are only $\sim20$\,ms apart, and the absence of CP in sub-pulse C2. Other bursts from this FRB source have shown a significant burst-to-burst RM variation but over much longer time scales of days to months \citep{Hilmarsson:2021_circular, Kumar:2022}. 

We extend our spectral modeling to fit the normalized residual LP spectra in C1 and the associated Stokes $V$ component using a GFR model \citep{Lower:2021}. We consider both the RM of C1 and C2 to get the de-Faraday rotated LP residuals for the sub-pulse C1. The limited bandwidth of the burst results in a strong degeneracy between the GRM and the spectral dependence, $\alpha$, leading to a poor reconstruction of the data using random samples drawn from the posteriors. We note that the median \textit{a-posteriori} model does appear to match the data well by eye. Nevertheless, we find that the sampled $\alpha$ for the two RM cases (C1 and C2) is in agreement within 1-$\sigma$ and 3-$\sigma$, respectively, with one arising for the wave propagation through a relativistic medium, i.e., $\alpha=3$ \citep{Melrose:1997}. Motivated by this consistency and to remove the strong covariance between parameters, we limit our analysis to a  relativistic case with $\alpha$ fixed to this theoretical value.

\begin{figure}
\centering
\includegraphics[width=1\columnwidth]{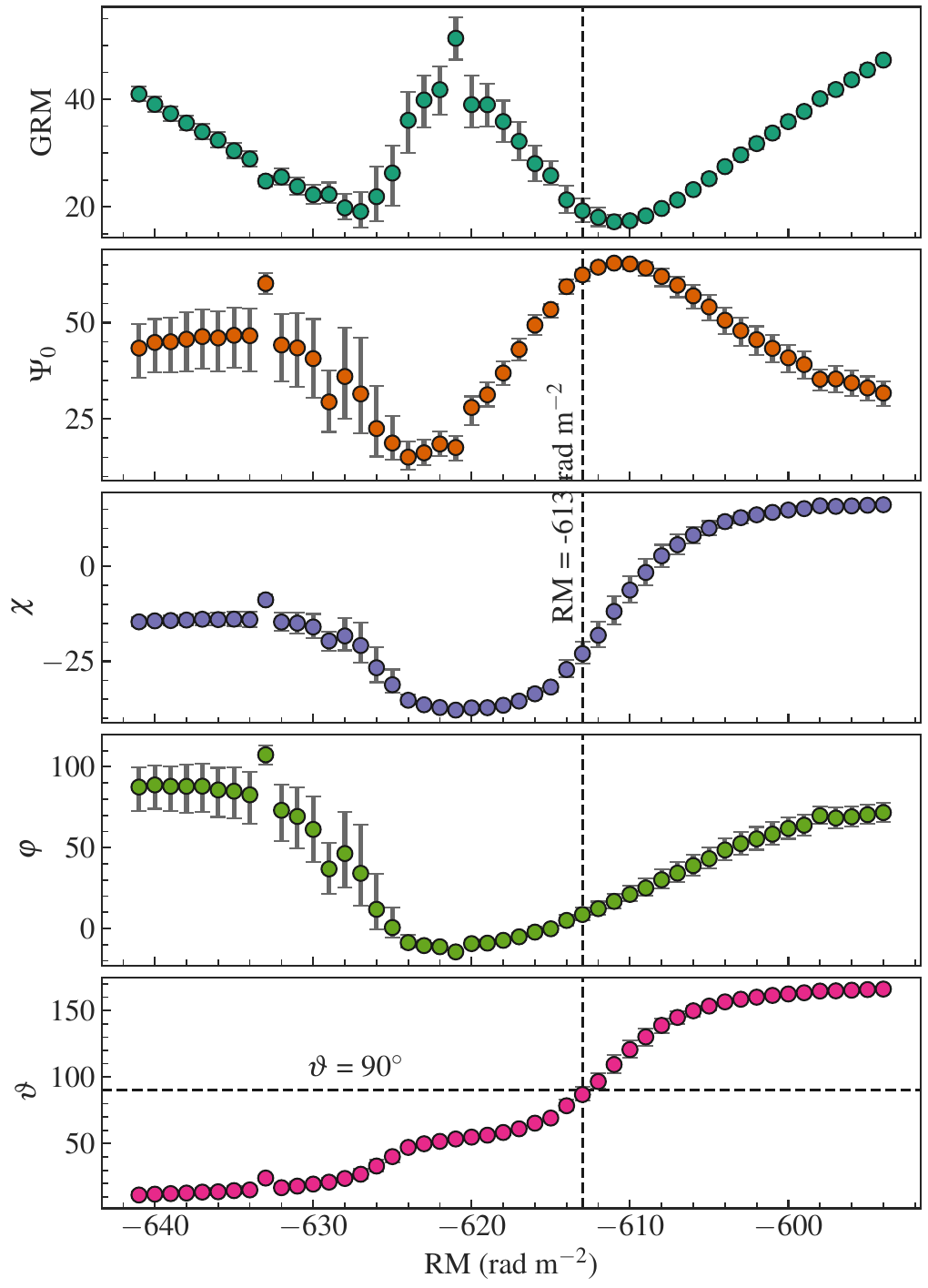} 
\caption{Best-fit GFR parameters for the FRB sub-pulse C1 Stokes spectra as a function of Faraday RM (for $\alpha=3$).}
\label{fig:rmgrid}
\end{figure}

\textit{Relativistic Rotation Measure ---} In this model with $\alpha=3$, the GFR is now described by an analogous relativistic rotation measure (RRM), which arises from the radiation propagation through a relativistic plasma. The polarization angle, in this case, varies with wavelength as $\Psi(\lambda) \propto \lambda^3\,\text{RRM}$ \citep{Kennett:1998}. To further probe the RM contamination to the net GFR, we now model the  Stokes spectrum of C1 on a grid of fixed RM values ($-594$ to $-642$~\RMunits). The obtained best-fit GFR parameters as a function of RM are shown in Figure \ref{fig:rmgrid}. We can clearly see a degeneracy between the RM and the GFR model parameters. The degeneracy highlights the complications associated with a GFR modeling when the net amount of regular FR along the LOS of the burst source is not known with significant precision. Nevertheless, in our case, we can utilize the sub-pulse C2 to calibrate for the regular FR in the intervening medium as the absence of CP undoubtedly indicates no Faraday conversion. Using the best-fit RM of sub-pulse C2, i.e., $-613$\RMunits~as the net FR contribution, we conduct parameter estimation and obtain RRM $=19\pm2$\RRMunits and $\vartheta=86^{+6}_{-5}$.

\begin{figure}
\centering
\includegraphics[width=1\columnwidth]{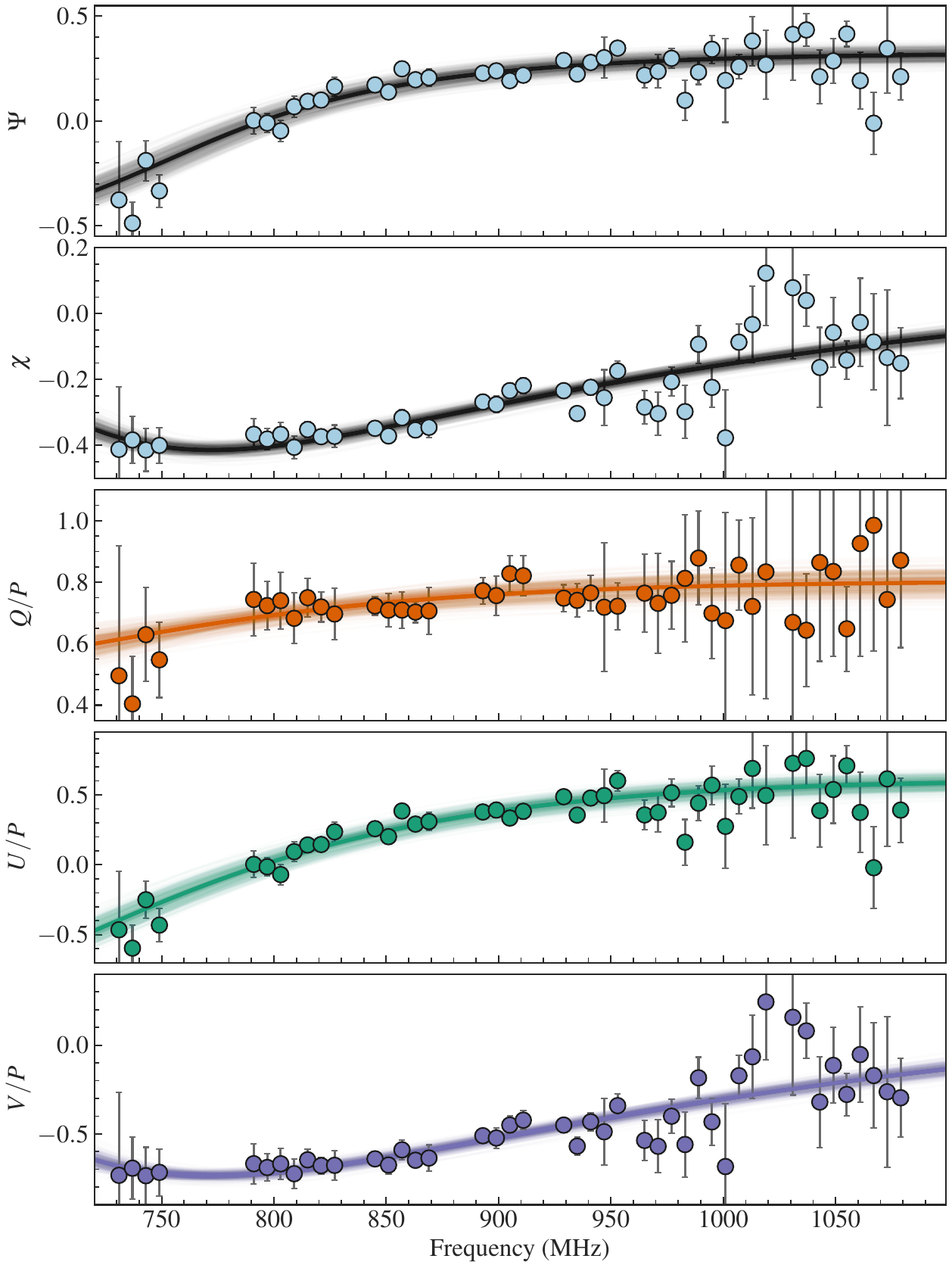} 
\caption{Fractional Stokes spectra of the FRB sub-pulse C1 after correcting for RM = $-613$~\RMunits. Thick solid lines indicate the median recovered GFR model for linearly polarized modes (relativistic case: $\alpha = 3$, $\vartheta = 90^{\circ}$), while the transparent traces indicate 350 random draws from the posterior distributions. The top two panels show the internal parameters (PA and ellipticity angle) calculated in the model. The unmodeled trend in the V/P panel (and as a result in the $\chi$ panel) around $\sim1.02$\,GHz is because of the increased level of radio frequency interference in these frequency channels, which affect the data quality.}
\label{fig:gfr_fit}
\end{figure}

\begin{figure}
\centering
\includegraphics[width=1\columnwidth]{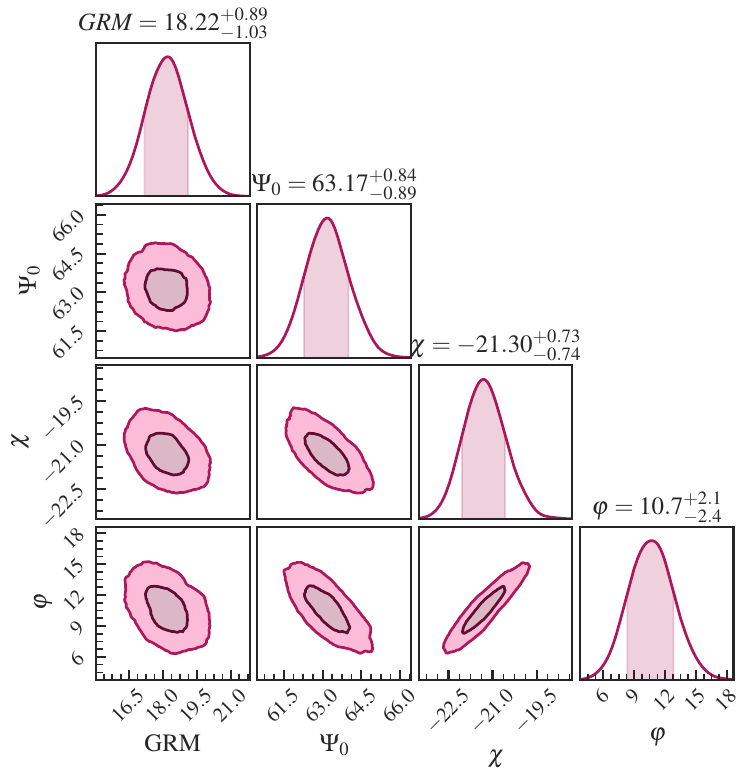} 
\caption{Posterior distributions of GFR modeling parameters for linearly polarized modes (relativistic case: $\alpha = 3$, $\vartheta = 90^{\circ}$). Top panels show the 1D marginalized distributions with the 68\% credible intervals as a shaded region. Other panels show the 2D distributions of these parameters, with the contours representing 68\% and 95\% credible intervals.}
\label{fig:gfr_linear}
\end{figure}

The best-fit parameter $\vartheta$ is consistent (within 1-$\sigma$) with $\vartheta=90^{\circ}$, a case where the ellipticity of the natural wave modes of the intervening relativistic plasma is linearly polarized. Additionally, $\vartheta$ displays a smooth variation with RM in the grid search (see Figure \ref{fig:rmgrid}). To simplify the plasma modes geometry, we run further estimation runs considering the propagating modes in the relativistic medium to be linearly polarized ($\vartheta = 90^{\circ}$). We now obtain an excellent fit to the Stokes spectra and well-constrained parameters with the RRM $=18\pm1$\RRMunits. The fractional stokes components, along with the median {\em a-posterior} recovered model, are shown in Figure \ref{fig:gfr_fit}. The posterior distribution of model parameters is in Figure \ref{fig:gfr_linear}.

We find that the Stokes spectra and the exceptionally high degree of CP of the sub-pulse C1 are consistent with arising from the radiation undergoing a GFR in a relativistic medium. The inferred modeling parameter $\vartheta$ indicates that the natural modes of such medium are linearly polarized, and the ultra-relativistic particles completely dominate. For linear wave modes, the birefringence results in a difference in the wave number $\Delta k \propto \omega^{-3}$ \citep{Sazonov:1969JETP, Melrose:1997PhRvE}. Assuming a power-law distribution of relativistic particles, the analogous relativistic Faraday RM can be expressed as \citep{Kennett:1998},
\begin{align}\label{eqn:rrm}
{\rm RRM} &= \nonumber \frac{e^4}{4 \pi^3 \varepsilon_0 m^3 c^4} \left(\frac{\beta-1}{\beta-2}\right) \int d\ell n_r B_\perp^2 \gamma \\
          &= 0.03\,L\,\langle n_r B_\perp^2 \gamma\rangle\,{\rm rad\,m^{-3}},
\end{align}
where $\beta$ is the power-law index, $\gamma$ is the minimum Lorentz factor, the magnetic field $B$ is in mG and $n_{r}$ is the number density of the relativistic particles in cm$^{-3}$. The final form of RRM in Equation~(\ref{eqn:rrm}) is independent of the form of the distribution function of particles \citep{Melrose:1997}. 

Assuming equipartition between the energy density of relativistic electrons and the magnetic energy density in the relativistic medium, Equation~(\ref{eqn:rrm}) can be approximated as RRM $\sim 0.001\,L\,B^4$\RRMunits \citep{Kennett:1998}. For an extended birefringent medium found in supernova remnants with a length scale of $L\sim1$pc, we obtain a net magnetic field strength of $\sim$12\,mG. This estimate is consistent with the typical magnetic field strengths observed in the pulsar wind nebulae \cite{Gaensler:2006}. However, it is unlikely for the effect of GFR to disappear within $\sim 20$\,ms (sub-pulse C2) on such an extended length scale. The size of the FRB emitting region, denoted as $r_{c}$, can be constrained by considering the light-crossing time. Using the temporal width of the sub-pulse C1 $\sim 18$\,ms, we get $r_{c}\lesssim5500$\,km. If the emission region and the GFR-inducing medium are co-located with $L\sim r_{c}$, we obtain a total magnetic field strength of $\sim$3.2\,G. In such a scenario, the time variability of plasma conditions will lead to variations in natural modes throughout the pulse duration. Consequently, the polarization vector will deviate from the expected behavior of $\alpha = 3$, and the detailed geometry of the emitting region should be taken into account \citep{Cho:2020}. Another possibility is if the emission region and the birefringent medium are spatially separated. However, without a valid model of the FRB progenitor and the emission physics, it is difficult to ascertain the location and size of the birefringent medium and remove the degeneracy with the magnetic field strength.

While the above analysis is valid for any medium dominated by relativistic particles, it is possible to have a plasma with an admixture of cold electrons. The plasma modes will not be completely linear in such a scenario, and the ratio of their number densities will determine the ellipticity \citep{Kennett:1998}. Nonetheless, in our case, the inferred value of $\vartheta$ and the conditions required for Equation~(\ref{eqn:rrm}) limits the scope to a medium dominated by relativistic particles. In addition to regular FR from the cold plasma along the LOS, it is also possible for FR (non-$\lambda^{2}$) to be induced by the hot plasma in the magnetosphere or the near wind zone due to an asymmetry in the relativistic particle distribution \citep{Wang:2011} or oblique propagation \citep{Lyutikov:2021}. We do not see any substantial non-$\lambda^{2}$ RM component in sub-pulse C2; we argue that any such contribution would be limited to sub-pulse C1. In that case, some unaccounted non-$\lambda^{2}$ RM contamination will be present in our inferred RRM. Given the low strength of the burst signal and limited bandwidth, it is beyond the scope of our method to account for these effects.

An ideal candidate for such a relativistic medium is a pair ($\mathrm{e^-e^+}$) plasma. A pair-plasma with equal distributions will also cancel out any regular FR or intrinsic synchrotron production of CP \citep{Kennett:1998}. Additionally, in such a plasma, the normal propagation modes are linearly polarized for all observable radio frequencies \citep{Zheleznyakov:2002}. Thus, the observed CP can entirely be attributed to the Faraday conversion or GFR in the plasma medium. Table~\ref{tab:table1} summarizes the various observable propagation effects in astrophysical plasmas. 

\begin{table}
\caption{Propagation effects in astrophysical plasma}
\begin{ruledtabular}
\begin{tabular}{lccc}
\multicolumn{1}{c}{\textrm{Effect}}&
\multicolumn{1}{c}{\textrm{Non-relativistic}}&
\multicolumn{1}{c}{\textrm{Relativistic ion}}&
\multicolumn{1}{c}{\textrm{Pair plasma}}\\
\colrule
\textrm{DM}  & $n_e L$  &  $n_r L / \gamma$  & $-$ \\
\textrm{RM}  & $n_e B_\parallel L$  & $\propto \lambda^{-1.5}$ \citep{Wang:2011} & 0 \\
\textrm{RRM} & $-$ & $n_r B_\perp^2 \gamma L $ & $n_r B_\perp^2 L $ \\
\end{tabular}
\end{ruledtabular}
\label{tab:table1}
\end{table}

\textit{DM Excess ---} The observed propagation delay as a function of frequency is used to quantify the amount of dispersion, DM (and consequently the electron column density) encountered by the radiation along the LOS. Primarily, this involves contributions from the cold plasma of the intervening medium
and thus only the non-relativistic case is considered. For a path length of $L$, the amount of dispersion is, ${\rm DM} = n_{e}L$, where $n_{e}$ is the number density of cold electrons. Thus, a measurement of RM and DM is used to estimate the average magnetic field strength along the LOS of an astrophysical signal (using Equation~(\ref{eqn:rm_fr}))
\begin{equation}\label{eqn:b_par}
\langle B_\parallel \rangle = 1.232 \left( \frac{{\rm RM}}{{\rm rad\,m^{-2}}}\right)\left( \frac{{\rm DM}}{{\rm pc\,cm^{-3}}}\right)^{-1} {\rm \upmu G}
\end{equation}
However, in a plasma entirely dominated by relativistic particles, the amount of dispersion is solely due to relativistic electrons ($n_{r}$) and can be given as (up to a first-order approximation), ${\rm DM_{rel}} = n_{r}L / 2\gamma$ \citep{Skilling:1971}. Thus, using an estimate of ${\rm DM_{rel}}$ and RRM, one can uniquely measure the average perpendicular magnetic field strength in such a plasma along the LOS (using Equation~(\ref{eqn:rrm}))
\begin{equation}\label{eqn:b_perp}
\langle B_\perp \gamma \rangle = 4.083 \left( \frac{{\rm RRM}}{{\rm rad\,m^{-3}}}\right)^{0.5}\left( \frac{{\rm DM_{rel}}}{{\rm pc\,cm^{-3}}}\right)^{-0.5} {\rm mG}
\end{equation}
Direct measurement of ${\rm DM_{rel}}$ and RRM is challenging as these are second-order observational effects, and one first needs to correctly account for the regular DM and RM due to the signal propagation in the intervening medium. With a precise estimate of the DM and RM along the LOS from long-term monitoring of the source, one can attribute the excess values in the measurements resulting from these relativistic effects in the GFR framework. A critical issue with this approach is in cases of repeating FRBs with substantial DM and RM variations but no correlation with the observed CP \citep{Hilmarsson:2021}. Such variations are most likely due to changes in the LOS through a dense and magnetized environment surrounding the FRB progenitors \citep{Michilli:2018, Niu:2021} and not due to propagation through a relativistic plasma \citep{Vedantham:2019}. 

However, we can further constrain these quantities using the weaker contemporaneous bursts. For the Parkes-detected repeat burst, an apparent excess DM of 5\DMunits~is inferred based on the complex frequency-time dynamic spectrum \citep{Kumar:2022}. We estimated an RRM of 18\RRMunits, considering the RM of sub-pulse C2 as the net FR contribution along the LOS. Furthermore, our GFR modeling with $\vartheta$ as a free parameter indicates it being consistent with $90^{\circ}$, implying that the relativistic plasma modes are linearly polarized. Assuming that the excess DM is due to propagation through this relativistic plasma, we estimate $\langle B_\perp \gamma \rangle = 8\pm1$\,mG. Consequently, for a medium with a minimum Lorentz factor, $\gamma > 10$, we can constrain the magnetic field strength to be $B_\perp < 1$\,mG. 

Considering the scenario where the emission region and the birefringent medium are co-located, and assuming that the relativistic plasma (with $L\sim5500$\,km) causes the excess dispersion, we would expect a plasma density of $n_{r} \sim 10^{11}$\,cm$^{-3}$. While this estimate is similar to the Goldreich-Julian plasma density limit in the magnetosphere of a young magnetar \citep{Goldreich:1969}, it is unlikely to cause the observed excess DM. Such a high plasma density would necessitate a net magnetic field strength on the order of $\sim10^{12}$\,G, assuming the magnetar spin-period $\sim$1\,s \citep{Lorimer:2004}. Moreover, such a strong magnetic field would likely suppress the plasma oscillations necessary to generate the observed excess DM. Nonetheless, we are not able to provide any substantial independent constraints on the net magnetic field with our GFR modeling. Whether such an extreme $B$-field orientation can exist in a relativistic medium is subject to further study. Hence, it is more probable that the FRB emission region and the birefringent medium are spatially separated. Determining the location of the GFR-inducing media along the line of sight remains a challenging task in our modeling efforts.

\textit{Implications ---} Repeating FRBs exhibit a variety of polarization properties. Prior to the discovery of the FRB\,20201124A, none of the repeating sources had shown evidence for CP. Studies in the context of the lack of observed CP and the likelihood of GFR in bursts from the most-studied repeating source, FRB\,20121102A \cite{Gruzinov:2019,Vedantham:2019}, placed limits on the magnetic geometry of its associated extreme magneto-ionic environment \citep{Michilli:2018}. Based on a complete polarized radiative transfer framework \citep{Huang:2011}, it has been argued that the persistent radio synchrotron nebula associated with FRB\,20121102A must be an admixture of highly relativistic and cold electrons ($\gamma \lesssim 3$) to explain the non-detection of CP \citep{Vedantham:2019}. Thus, measurements of CP in repeating FRBs provides a means to probe the low-energy end ($\gamma \lesssim 100$) of the relativistic electron distribution, which is otherwise not observable with standard methods \citep{Vedantham:2019}. 

Unlike relatively stable polarization features in most repeating FRB sources, the FRB\,20201124As source shows significant LP and CP changes across its repeat bursts. The presence of CP is also not uniform in bursts even on a scale of a few milliseconds \citep{Hilmarsson:2021_circular, Kumar:2022}, suggesting a time-variable mechanism responsible. Additionally, similar large PA swings of $\sim 50^{\circ}$, as seen across sub-pulse C1 in Figure \ref{fig:burst}, are not seen in either sub-pulse C2 or subsequently detected bursts. The downward curve of the PA swing in C1, not very frequently seen in FRBs, is somewhat reminiscent of magnetars and radio pulsars. Such PA variation is usually interpreted in terms of structural changes around the progenitor magnetosphere during the activity cycle and the burst emission originating from different regions \citep{Noutsos:2009, Dai:2019, Ilie:2019}. However, an important distinction is that the observed PA variation is averaged over many rotations in the case of pulsars and magnetars and thus can have a somewhat different distribution due to the presence of orthogonal polarization modes \citep{Backer:1976}. The non-detection of GFR and similar frequency-dependent CP in other bursts from this source indicates the rarity of the necessary magnetic field alignments and the plasma conditions along the LOS and sets the FRB20201124A source as an extreme case in the FRB population.

GFR and relativistic plasma effects can potentially be further probed in similar bright bursts from FRB\,20201124A and other FRB sources. Detection of a positive correlation between the Faraday RM variations in subsequent bursts with the presence of CP would be compelling evidence for the GFR. If the CP originates in the magnetosphere, we would also expect a correlation between the degree of observed CP and the PA variation across the pulse profile \citep{Han:1998}. Detection of excess dispersion along with GFR in these bursts can directly probe the perpendicular component of the magnetic field strength in the plasma \citep{Li:2019}. However, it is subject to further study if a relativistic plasma can generate such observational changes in the burst DM.

Repeating FRBs with high activity rates provides a unique way to probe the relativistic plasma. Detection of CP provides a powerful diagnostic of magnetic field geometry and particle densities in the vicinity of the progenitor compared to LP. We showed that the observed high levels of CP in a repeat burst from FRB\,202021124A are consistent with being generated via a generalized form of Faraday rotation. We showed that the properties of the intervening medium responsible for the GFR are consistent with a plasma dominated by relativistic particles, possibly a pair-plasma. Such a medium can also explain the apparent DM excess inferred from the burst spectra, which can then be used to estimate the perpendicular B-field. We discussed the complications associated with GFR modeling and the significance of having a precise estimate of RM and DM along the LOS to estimate the second-order relativistic effects inducing GFR. Long-term monitoring of these spectro-polarimetric effects across the population of repeating FRBs could enable detailed studies of the plasma composition and magnetic topology of the immediate environment of the progenitor. This may, in turn, allow us to place constraints on the types of objects that produce repeating FRBs.

\medskip

\textit{Acknowledgments ---}
The Parkes radio telescope (\textit{Murriyang}) is part of the Australia Telescope National Facility (\url{https://ror.org/05qajvd42}) which is funded by the Australian Government for operation as a National Facility managed by CSIRO. We acknowledge the Wiradjuri people as the traditional owners of the Observatory site.
This work was performed on the OzSTAR national facility at Swinburne University of Technology. The OzSTAR program receives funding in part from the Astronomy National Collaborative Research Infrastructure Strategy (NCRIS) allocation provided by the Australian Government.
P.K. acknowledge support through the Australian Research Council (ARC) grant FL150100148. R.M.S. acknowledges support through ARC grants  DP180100857 and FT190100155.
J.X.P. as a member of the Fast and Fortunate for FRB Follow-up team, acknowledges support from NSF grants AST-1911140 and AST-1910471.

\bibliographystyle{apsrev4-2}
\bibliography{references}


\end{document}